\documentstyle[twocolumn,aps,prl]{revtex}

\begin{document}

\draft

\title{Finite-element theory of transport in ferromagnet-normal metal systems}

\author{Arne Brataas$^{*}$, Yu. V. Nazarov and Gerrit E. W. Bauer}

\address{Department of Applied Physics and Delft Institute of
Microelectronics and Submicrontechnology,\\ Delft University of
Technology, Lorentzweg 1, 2628 CJ Delft, The Netherlands}

\date{\today} \maketitle

\begin{abstract}

We formulate a theory of spin dependent transport of an electronic
circuit involving ferromagnetic elements with non-collinear
magnetizations which is based on the conservation of spin and
charge current. The theory considerably simplifies the
calculation of the transport properties of complicated
ferromagnet-normal metal systems. We illustrate the theory by
considering a novel three terminal device.

\end{abstract}

\pacs{72.10.Bg,72.10.-d,75.70.Pa}

Electron transport in hybrid systems involving ferromagnetic and
normal metals has been shown to exhibit new phenomena due to the
interplay between spin and charge. The giant magnetoresistance
(GMR) effect in metallic magnetic multilayers is a result of spin
dependent scattering\cite{Levy94:367}. The manganese oxides
exhibit a colossal magnetoresistance \cite{Moreo99:2034} due to a
ferromagnetic phase transition. The dependence of the current on
the relative angle between the magnetization directions has been
reported in transport through tunnel junctions between
ferromagnetic reservoirs \cite{Moodera96:4724}. Transport involving
ferromagnets with non- collinear magnetizations has also been
studied theoretically in Ref.\ \cite{Slon89}

Johnson and Silsbee demonstrated that spin dependent effects are also
important in systems with more than two terminals
\cite{Johnson85:1790}. Their ferromagnetic-normal-ferromagnetic
(F-N-F) device manifests a transistor effect that depends on the
relative orientation of the magnetization directions.  Recently
another three terminal spin electronics device was realized; a
ferromagnetic single-electron transistor\cite{Ono96:3449}. In this
case the current depends on the relative orientation of the
magnetization of the source, the island and the drain and of the
electrostatic potential of the island tuned by a gate voltage
\cite{Brataas99:93}.

These examples illustrate that devices with ferromagnetic order
deserve a thorough theoretical investigation. Inspired by the circuit
theory of Andreev reflection \cite{Nazarov94:1420} we present a
finite-element theory for transport in hybrid ferromagnetic-normal
metal systems based on the conservation of charge and spin current. We
demonstrate that spin-transport can be understood in terms of 4
generalized conductances for each contact between a ferromagnet and a
normal metal. The relations between these conductance parameters and
the microscopic details of the contacts are derived and calculated for
diffuse, tunnel and ballistic contacts.  Finally, we illustrate the
theory by computing the current through a novel 3-terminal device.

Let us first explain the basic idea of the finite-element theory of
spin-transport. The system can be divided into (normal or
ferromagnetic) ``nodes", where each node is characterized by the
appropriate generalization of the distribution function, {\em viz.} a
$2\times 2$ distribution matrix in spin space. The nodes are connected
to each other and to the reservoirs by ``contacts" which limit the
total conductance but are arbitrary otherwise. The charge and spin
current through the contacts is related to the distribution matrices
of the adjacent nodes. Provided these relations are known, we can
solve for the $2 \times 2$ distribution matrices in the nodes under
the constraint of conservation of spin and charge current in each node
and thus determine the transport properties of the system. These
macroscopic relations for each contact can be found in terms of the
microscopic scattering matrices in the spirit of the
Landauer-B\"{u}ttiker formalism \cite{Buttiker86:1761}. The scattering
matrices can be calculated using different models like a two-spin band
model or realistic band-structures and for various contacts, {\em
e.g.} ballistic or diffuse wires or tunnel junctions. Phase coherent
scattering as in a resonant tunneling devices and effects like the
Coulomb blockade can be included in principle by calling the double
barrier a "contact" with complex scattering properties, but these
complications will be disregarded in the following.

The device depicted in Fig.\ \ref{f:cir} will serve to illustrate our
approach. Several contacts attach a normal metal node to
(ferromagnetic or normal) metallic reservoirs. We assume that the
resistances of the contacts are much larger than the resistance of the
node.  This is fulfilled when the area of the contact is sufficiently
smaller than the cross-section of the node or when the contacts are in
the tunneling regime. The current through the system and the
distribution matrix in the node are determined by the properties of
the contacts. The reservoirs are supposed to be large and in local
equilibrium with a chemical potential $\mu_{\alpha}$, where the
subscript $\alpha$ labels the reservoirs. The energy dependent
distribution matrix in the (ferromagnetic) reservoir is then diagonal
in spin-space
$\hat{f}_{\alpha}^F(\epsilon)=\hat{1}f(\epsilon,\mu_{\alpha})$, where
hat ($\hat{}$) denotes a $2\times2$ matrix in spin space, $\hat{1}$ is
the unit matrix and $f(\epsilon,\mu_{\alpha})$ is the the Fermi-Dirac
distribution function. The direction of the magnetization is denoted
by the unit vector ${\bf m}_{\alpha}$.  When the chemical potentials
of the reservoirs are not identical, the normal metal node is not in
equilibrium and there can be a spin-accumulation on the normal metal
node. The distribution is therefore represented by a $2\times2$ matrix
in spin-space, $\hat{f}^N(\epsilon)$ which allows a spin accumulation
with arbitrary direction of the spins. The normal metal node is
considered to be large and chaotic either because of impurity
scattering inside the node or because of scattering at irregularities
of its boundary. The distribution matrix inside the node is therefore
isotropic in momentum space and depends only on the energy of the
particle.

The current through a contact is determined by its scattering matrix,
the Fermi-Dirac distribution function of the adjacent ferromagnetic
reservoir and the $2 \times 2 $ non-equilibrium distribution matrix in
the normal node. The current is evaluated close to the contact on the
normal side. The $2 \times 2$ current in spin-space per energy
interval at energy $\epsilon$ leaving the node is
\begin{equation}
\frac{h}{e^2} \hat{i} = 
\sum_{nm} \left[ \hat{r}^{nm}
\hat{f}^{N}(\hat{r}^{nm})^{\ast} +
\hat{t}^{nm}\hat{f}^{F}(\hat{t}^{nm})^{\ast} \right] - M \hat{f}^{N}
\label{icon}
\end{equation}
where $M$ is the number of propagating channels,
$\hat{r}^{nm}(\epsilon)$ is the reflection matrix for an electron
coming from the normal metal in mode $m$ being reflected to mode $n$
and $\hat{t}^{nm}(\epsilon)$ is the transmission matrix for an
electron from the ferromagnet in mode $m$ transmitted to the normal
metal in mode $n$. The total current is obtained by integrating over
the energies, $\hat{I}=\int d\epsilon \hat{i}(\epsilon)$. The current
in the contact is thus completely determined by the scattering matrix
of the contact, and the distribution matrices.

The $2 \times 2$ non-equilibrium distribution matrix in the node in
the stationary state is uniquely determined by current conservation
\begin{equation}
\sum_{\alpha} \hat{i}_{\alpha} = \left( \frac{\partial
\hat{f}^N}{\partial t} \right)_{\text{rel}} \, , \label{curcons}
\end{equation}
where $\alpha$ labels different contacts and the term on the right
hand side describes spin relaxation in the normal node. The right hand
side of Eq.\ (\ref{curcons}) can be set to zero when the spin current
in the node is conserved, {\em i.e.} when an electron spends much less
time on the node than the spin-flip relaxation time
$\tau_{\text{sf}}$. If the size of the node in the transport direction
is smaller than the spin-flip diffusion length $l_{\text{sf}} =
\sqrt{D \tau_{\text{sf}}}$, where $D$ is the diffusion coefficient
then the spin relaxation in the node can be introduced as $(\partial
\hat{f}^N / \partial t)_{\text{rel}} = (\hat{1}\text{Tr}(\hat{f}^N)/2
- \hat{f}^N)/\tau_{\text{sf}}$.  If the size of the node in the
transport direction is larger than $l_{\text{sf}}$ the simplest
finite-element transport theory fails and we have to use a more
complicated description with a spatially dependent spin distribution
function\cite{Huertas99}. Eq.\ (\ref{curcons}) gives the $2 \times 2$
distribution matrix of the node in terms of Fermi-Diract distribution
functions of the reservoirs. These distribution functions are
determined by voltages of the reservoirs. Those voltages are either
set by voltage sources or determined by conventional circuit theory.

We will now demonstrate that the relation (\ref{icon}) between the
current and the distributions has a general macroscopic
form. Spin-flip processes in the contacts are disregarded, so that the
reflection matrix for an incoming electron from the normal metal can
be written as $\hat{r}^{nm}=\sum_s\hat{u}^s r_s^{nm}$, where
$s=\uparrow,\downarrow$, $r_{s}^{nm}$ are the spin dependent
reflection coefficients in the basis where the spin quantization axis
is parallel to the magnetization in the ferromagnet,
$\hat{u}^{\uparrow}= (\hat{1} + \hat{\text{\boldmath $\sigma$}} \cdot
{\bf m})/2$,
$\hat{u}^{\downarrow} = (\hat{1} - \hat{\text{\boldmath $\sigma$}}
\cdot {\bf m})/2$ and $\hat{\text{\boldmath $\sigma$}}$ is a vector of
Pauli matrices.  Similarly for the transmission matrix
$ \hat{t}^{nm} (\hat{t}^{nm})^{\ast}=
\sum_{s} \hat{u}^s |t^{nm}_{s}|^2 \, , $
where $t_{s}^{nm}$ are the spin dependent transmission coefficients.
Using the unitarity of the scattering matrix, we find that the general
form of the relation (\ref{icon}) reads
\begin{eqnarray}
\hat{i} & = &
G^{\uparrow} \hat{u}^{\uparrow} \left( \hat{f}^F - \hat{f}^N \right)
\hat{u}^{\uparrow}
+ G^{\downarrow} \hat{u}^{\downarrow} \left( \hat{f}^F - \hat{f}^N
\right) \hat{u}^{\downarrow} \nonumber \\
&& - G^{\uparrow \downarrow} \hat{u}^{\uparrow} \hat{f}^N
\hat{u}^{\downarrow}
- (G^{\uparrow \downarrow})^{\ast} \hat{u}^{\downarrow} \hat{f}^N
\hat{u}^{\uparrow} \, ,
\label{curgen}
\end{eqnarray}
where we have introduced the spin dependent conductances $G^s$
\begin{eqnarray}
G^{s} = \frac{e^2}{h} \left[ M-\sum_{nm} |r^{nm}_{s}|^2 \right] =
\frac{e^2}{h} \sum_{nm} |t^{nm}_s|^2 \label{Gspin}
\end{eqnarray}
and the mixing conductance
\begin{equation}
G^{\uparrow \downarrow}=\frac{e^2}{h} \left[ M-\sum_{nm}
r_{\uparrow}^{nm} (r_{\downarrow}^{nm})^{\ast} \right] \, .
\label{Gmixing}
\end{equation}
We thus see that the relation between the current through a contact
and the distribution in the ferromagnetic reservoir and the normal
metal node is determined by 4 conductances, the two real spin
conductances ($G^{\uparrow}$, $G^{\downarrow}$) and the real and
imaginary parts of the mixing conductance $G^{\uparrow
\downarrow}$. These contact-specific parameters can be obtained by
microscopic theory or from experiments. The spin conductances
$G^{\uparrow}$ and $G^{\downarrow}$ have been used in descriptions of
spin-transport for a long time \cite{Levy94:367}. The {\em mixing
conductance} is a new concept which is relevant for transport between
non-collinear ferromagnets. The mixing conductance rotates spins
around the magnetization axis of the ferromagnet. Note that although
the mixing conductance is a complex number the $2\times 2$ current in
spin-space is hermitian and consequently the current and the
spin-current in an arbitrary direction given by Eq.\ (\ref{curgen})
are real numbers. Generally we can show that $\text{Re} G^{\uparrow
\downarrow} \ge (G^{\uparrow}+G^{\downarrow})/2$. Below we present
explicit results for the conductances when the contacts are in the
diffuse, tunneling and ballistic regimes.

For a diffuse contact Eq.\ (\ref{curgen}) can quite generally be found
by the Green function technique developed in Ref.\
\cite{Nazarov94:134}. Here we use a much simpler approach based on the
diffusion equation. On the normal metal side of the contact the
boundary condition to the diffusion equation is set by the
distribution matrix in the node $\hat{f}^N$. On the ferromagnet side
of the contact the boundary condition is set by the equilibrium
distribution function in the reservoir $f^F \hat{1}$.  In a
ferromagnetic metal transport of spins non-collinear to the local
magnetization leads to a relaxation of the spins since electrons with
different spins are not coherent. This causes an additional
resistance, which as other interface related excess resistances, is
assumed to be small compared to the diffuse bulk
resistance. Sufficiently far from the ferromagnetic-normal metal
interface the distribution function of the electronic states in the
ferromagnet can always be represented by two components.  Only the
spin-current parallel to the magnetization of the ferromagnet is
conserved. We denote the cross-section of the contact $A$, the length
of the ferromagnetic part of the contact $L^F$, the length of the
normal part of the contact $L^N$, the (spin dependent) resistitivity
in the ferromagnet $\rho^{Fs}$, the resistivity in the normal metal
$\rho^N$, so that the (spin dependent) conductance of the
ferromagnetic part of the contact is $G^{DFs}=A/(\rho^{Fs}L^F)$ and
the conductance of the normal part of the contact is $G^{DN}=A/(\rho^N
L^N)$. Solving the diffusion equation $\nabla^2 \hat{f}=0$ on the
normal and ferromagnetic side with the boundary conditions above, we
find the current through a diffuse contact:
\begin{eqnarray}
\hat{i}^D & = &
G^{D\uparrow} \hat{u}^{\uparrow} \left( \hat{f}^F - \hat{f}^N \right)
\hat{u}^{\uparrow}
+ G^{D\downarrow} \hat{u}^{\downarrow}
\left(\hat{f}^F - \hat{f}^N \right) \hat{u}^{\downarrow}
\nonumber \\
&& - G^{DN} \left( \hat{u}^{\uparrow} \hat{f}^N
\hat{u}^{\downarrow}
+ \hat{u}^{\downarrow} \hat{f}^N \hat{u}^{\uparrow}
\right) \, ,
\label{curdiff}
\end{eqnarray}
where the total spin dependent conductance is
$1/G^{Ds}=1/G^{DFs}+1/G^{DN}$. This result can be understood as a
specific case of the generic Eq.\ (\ref{curgen}) with
$G^{\uparrow}=G^{D\uparrow}$, $G^{\downarrow}=G^{D\downarrow}$,
and $G^{\uparrow \downarrow}=G^{DN}$. The mixing conductance in
the diffuse limit therefore depends on the conductance of the
normal part of the contact only, which is a consequence of the
relaxation of spins non-collinear to the magnetizations direction
in the ferromagnet.

For a ballistic contact, we use a simple semiclassical model proposed
in Ref.\ \cite{Jong95:1657}. In this model the channels are either
completely reflected or transmitted, with $N^{\uparrow}$ and
$N^{\downarrow}$ being the number of transmitted channels for
different spin directions.  Substituting this in (\ref{icon}) we find
that the spin conductance $G^{B\uparrow}=(e^2/h)N^{\uparrow}$,
$G^{B\downarrow}=(e^2/h)N^{\downarrow}$ and the mixed conductance is
determined by the lowest number of reflected channels,
$G^{B\uparrow\downarrow}=\text{max} (G^{B\uparrow},G^{B\downarrow})$ and
is real. 

For a tunneling contact we can expand Eq.\ (\ref{icon}) in terms of the small transmission. We find that
$\text{Re} G^{T\uparrow \downarrow} =
(G^{T\uparrow}+G^{T\downarrow})/2$, where $G^{T\uparrow}$ and
$G^{T\downarrow}$ are the tunneling conductances. The imaginary part of $G^{T\uparrow
\downarrow}$ can be shown to be of the same order of magnitude as $G^{T\uparrow}$ and
$G^{T\downarrow}$ but it is not universal.

We will now illustrate the theory by computing the current through
the 3-terminal device shown in Fig.\ \ref{f:two}. A normal metal
node (N) is connected to 3 ferromagnetic reservoirs (F1, F2 and
F3) by arbitrary contacts parameterized by our spin-conductances.
A source-drain bias voltage $V$ applied between reservoir 1 and 2
causes an electric current $I$ between the same reservoirs. The
charge flow into reservoir 3 is adjusted to zero by the chemical
potential $\mu_3$. Still, the magnetization direction ${\bf m}_3$
influences the current between reservoir 1 and 2. We assume that
spin relaxation in the normal node can be disregarded so that the
right hand side of (\ref{curcons}) is set to zero. Furthermore, we
assume that the voltage bias $V$ is sufficiently small so that the
energy dependence of the transmission (reflection) coefficients
can be disregarded. To further simplify the discussions the
contacts 1 and 2 are taken to be identical,
$G_{1}^{\uparrow}=G_{2}^{\uparrow} \equiv G^{\uparrow}$,
$G_{1}^{\downarrow}=G_{2}^{\downarrow} \equiv G^{\downarrow}$ and
$G_{1}^{\uparrow \downarrow}=G_{2}^{\uparrow \downarrow} \equiv
G^{\uparrow \downarrow}$. Contact 3 is characterized by the
conductances $G_{3}^{\uparrow}$, $G_3^{\downarrow}$ and
$G_3^{\uparrow \downarrow}$. We find the distribution in the
normal node by solving the 4 linear Eqs.\ (\ref{curcons}). The
current through the contact between reservoir 1 (2) and the node
is obtained by inserting the resulting distribution for the normal
node into Eq.\ (\ref{curgen}).

When the magnetizations in reservoir 1 and 2 are parallel there is no
spin-accumulation since contacts 1 and 2 are symmetric and
consequently ferromagnet 3 does not affect the transport
properties. The current is then simply a result of two total
conductances $G=G^{\uparrow} +G^{\downarrow}$ in series, $I=G
V/2$. The influence of ferromagnet 3 is strongest when there is a
significant spin accumulation in the normal metal node, and in the
following the magnetizations of the source and drain reservoirs are
antiparallel, ${\bf m}_1 \cdot {\bf m}_2 =-1$.  We denote the relative
angle between the magnetization in reservoir 3 and reservoir 1
(reservoir 2) $\theta_3$ ($\pi-\theta_3$). The current is an even
function of $\theta_3$ and symmetric with respect to $\theta_3
\rightarrow \pi-\theta_3$ as a result of the symmetry of the device,
e.g. the current when the magnetizations in reservoir 1 and 3 are
parallel equals the current when the magnetizations in reservoir 1 and
3 are antiparallel. Due to the finite mixing conductance at
non-collinear magnetization the third contact acts as a drain for the
spin-accumulation in the node, thus allowing a larger charge current
between reservoir 1 and 2. The relative increase of the current due to
the reduced spin-accumulation $\Delta_3(\theta_3) =
[I(\theta_3)-I(\theta_3=0)]/I(\theta_3=0)$, is plotted in Fig.\
\ref{f:three} as a function of $\theta_3$. The maximum of $\Delta_3$
is achieved at $\theta_3=\pi/2$ ($\theta_3=3\pi/2$) and equals
($\text{Im}G^{\uparrow \downarrow}_3=0$)
\begin{equation}
\Delta_3 =P^2 \frac{2GG_3} {2G+G_3 \eta_3} \cdot
\frac{\eta_3-1+P_3^2} {2G(1-P^2)+G_3(1-P_3^2)} \,
\end{equation}
introducing the total conductance of the contact
$G_i=G_i^{\uparrow}+G_i^{\downarrow}$, the polarization of the contact
$P_i=(G_i^{\uparrow}-G_i^{\downarrow})/(G_i^{\uparrow}+G_i^{\downarrow})$
and the relative mixing conductance $\eta=2G_i^{\uparrow
\downarrow}/(G_i^{\uparrow}+G_i^{\downarrow})$. The influence of the
direction of the magnetization of the reservoir 3 increases with
increasing polarization $P$ and increasing relative mixing conductance
$\eta_3$ and reaches its maximum when the total conductances are of
the same order $G_3 \sim G$. Note that the physics of this three
terminal device is very different from that of Johnson's spin
transistor\cite{Johnson85:1790}; the latter operates with collinear
magnetizations of two ferromagnetic contacts whereas the third may be
normal.

In conclusion we have proposed a finite-element transport theory for
spin transport in mesoscopic systems. In the presence of ferromagnetic
order a contact can be described by 4 conductance parameters which we
obtained explicitly for diffuse, ballistic and tunnel contacts. We
have applied the theory to a novel three terminal device with
arbitrary contacts.

This work is part of the research program for the ``Stichting voor
Fundamenteel Onderzoek der Materie'' (FOM), which is financially
supported by the ''Nederlandse Organisatie voor Wetenschappelijk
Onderzoek'' (NWO). We acknowledge benefits from the TMR Research
Network on ``Interface Magnetism'' under contract No. FMRX-CT96-0089
(DG12-MIHT) and support from the NEDO joint research program
(NTDP-98). We acknowledge stimulating discussions with Wolfgang Belzig,
Daniel Huertas Hernando and Junichiro Inoue.

\begin{figure}
\caption{A normal node connected to ferromagnetic reservoirs
characterized by the chemical potentials $\mu_{\alpha}$ and the
magnetization vector ${\bf m}_{\alpha}$. The distribution matrix in
the normal node $\hat{f}^N$ can be found from the Kirchoff rules for
the spin currents $\hat{i}_{\alpha}$.}
\label{f:cir}
\end{figure}

\begin{figure}
\caption{The 3-terminal device where a normal metal node is
connected to ferromagnetic reservoirs. An applied bias causes a
source-drain current between F1 and F2. The charge current into F3
is adjusted to vanish by ${\mu}_3$. The magnetization direction of
ferromagnet F3 controls the current. } \label{f:two}
\end{figure}

\begin{figure}
\caption{The current $\text{Tr}(i_1)=-\text{Tr}(i_2)$ in the three
terminal device as a function of the magnetization direction
$\theta_3$. The straight, dashed and dotted lines correspond to
the relative mixing conductance $\text{Re}\eta_2=2$, $5$, and $10$
respectively. The other parameters were set to $P=0.4$, $P_3=0.1$,
$G=G_3$,$\text{Re} \eta_3=1.0$, $\text{Im} \eta = 0 =\text{Im}
\eta_3$.} \label{f:three}
\end{figure}

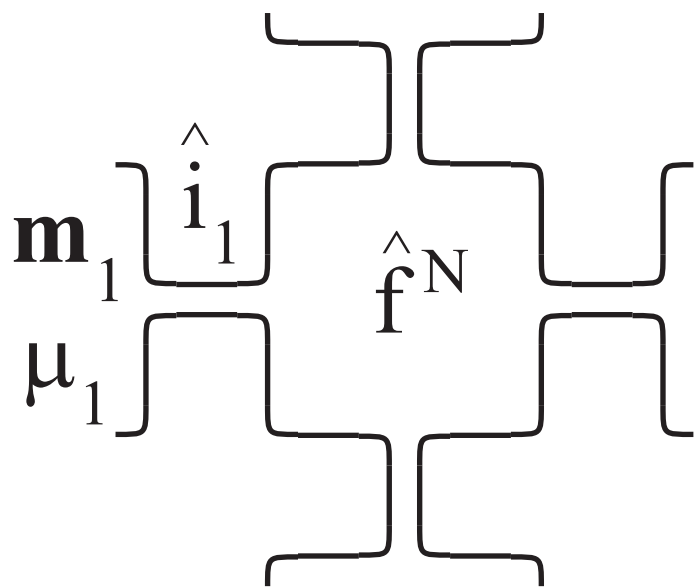
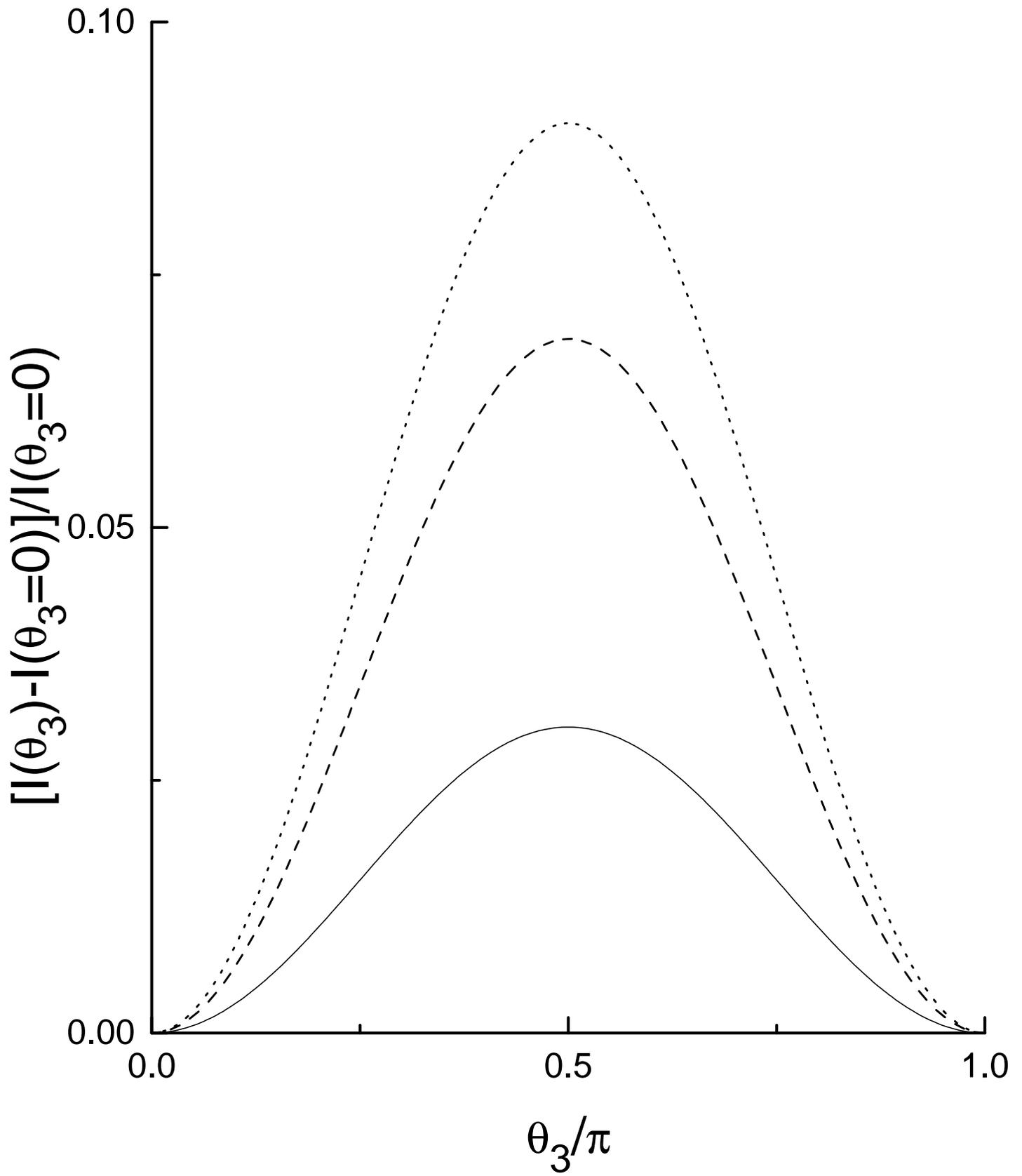
\end{document}